\documentclass[]{aa}
\usepackage{natbib} 
\bibpunct{(}{)}{;}{a}{}{,} 
\usepackage{graphicx}
\usepackage{txfonts}
\def\inte{{\em INTEGRAL}}
\def\xmm{{\em XMM-Newton}}

\def\chan{{\em Chandra}}
\def\asca{{\em ASCA}}

\def\swift{{\em Swift}}
\def\rosat{{\em ROSAT}}
\def\einst{{\em Einstein}}
\def \inte {{$INTEGRAL$}}
\def \xmm {{\em XMM--Newton}}

\def \ferg {\mbox{erg cm$^{-2}$ s$^{-1}$}}
\def \hcm {\hbox {\ifmmode $ atom cm$^{-2}\else atom cm$^{-2}$\fi}}

\def \arcsec {\hbox{$^{\prime\prime}$}}

\def \apjl {ApJL}
\def \apjs {ApJS}\def \aap {A\&A}

\def \igrA{IGR\,J11014-6103} 
\def\j11{IGR~J11014-6103}
\def\inte{{\em INTEGRAL}}
\def\xmm{{\em XMM-Newton}}

\def\rosat{{\em ROSAT}}
\def\chan{{\em Chandra}}
\def\asca{{\em ASCA}}

\def\swift{{\em Swift}}
\def\einst{{\em Einstein}}
\def\fermi{{\em Fermi}}
\def\egret{{\em EGRET}}
\def\SNR{{MSH 11-61A}}

\begin{document}
   \title{ \igrA:\ a newly discovered pulsar wind nebula?}

\author{L. Pavan 
  \inst{1}
  \and 
  E. Bozzo 
  \inst{1} 
  \and  
  G. P{\"u}hlhofer \inst{2}
  \and
  C. Ferrigno  
  \inst{1}  
  \and	
  M. Balbo 
  \inst{1}
  \and 
  R. Walter 
  \inst{1}		
}

\authorrunning{L. Pavan et al.}
  \titlerunning{ \igrA,\ a new PWN?}
  \offprints{lucia.pavan@unige.ch}

\institute{ISDC data center for astrophysics of the University of Geneva
	 chemin d'\'Ecogia, 16 1290 Versoix Switzerland. 
	 \and
	  Institut f{\"u}r Astronomie und Astrophysik, Kepler Center for Astro and Particle Physics, 
          Eberhard-Karls-Universit{\"a}t, Sand 1, 72076 T{\"u}bingen, Germany
	}

 \abstract{\igrA\ is one of the still unidentified hard X-ray \inte\ sources, 
 reported for the first time in the 4th IBIS/ISGRI catalog.}  
 {We investigated the nature of \igrA\ by carrying out a multiwavelength 
 analysis of the available archival observations performed in the direction of the source.}  
 {We present first the results of the timing and spectral analysis of all the X-ray 
 observations of \igrA\ carried out with \rosat, \asca, \einst, \swift, and \xmm,\ and then 
 use them to search for possible counterparts to the source in the optical, infra-red,
 radio and $\gamma$-ray domain.} 
 {Our analysis revealed that \igrA\ is comprised of three different X-ray emitting 
 regions: a point-like source, an extended object and a cometary-like ``tail'' ($\sim$4~arcmin).   
 A possible radio counterpart positionally coincident with the source was also identified.} 
 {Based on these results, we suggest that the emission from \igrA\ is generated by 
a pulsar wind nebula produced by a high-velocity pulsar.
\igrA\ might be the first of these systems detected with \inte\ IBIS/ISGRI.}
   

   \date{Received 2011 May 31; accepted 2011 July 09}

   \maketitle

\section{Introduction}
\label{sec:intro} 

The IBIS/ISGRI telescope \citep{ubertini03} onboard \inte\ \citep{winkler03} 
has been operating since seven years from now, and the latest available catalog 
of sources detected with this telescope contains more that 700 objects \citep{bird10}. 
A large fraction of them belongs to the class of X-ray binaries (26\%) and active 
galactic nuclei (35\%); a relatively small fraction is represented by cataclysmic 
variables (5\%), and another 4\% comprises supernova remnants, 
globular clusters and soft $\gamma$-ray repeaters.    
About 30\% of the IBIS/ISGRI sources are still unclassified 
\citep[see e.g.][for a recent review]{chaty10}. 

In this paper, we investigate the nature of the still unidentified \inte\ 
source \igrA.\ This source was reported 
for the first time in the 4th IBIS/ISGRI catalog \citep{bird10}, and was 
detected in the IBIS/ISGRI mosaic at a significance level of 5.4~$\sigma$ (20-100~keV). 
The best determined \inte\ position of \igrA\ is at \mbox{$RA$=165.341~deg,} \mbox{$Dec$=-61.056~deg} 
(J2000), with an associated uncertainty of 4.3~arcmin (90\% c.l.). 
The averaged source fluxes in the 20-40~keV and 40-100~keV energy band are  
0.4$\pm$0.1~mCrab and 0.6$\pm$0.2~mCrab, respectively\footnote{1~mCrab 
corresponds to 7.57$\times$10$^{-12}$~\ferg\  
in the 20-40 keV energy band and to 9.42$\times$10$^{-12}$~\ferg\  in the 
40-100~keV energy band.}. 
The light curve of the source, obtained
from the online tool {\sc Heavens}\footnote{http://www.isdc.unige.ch/heavens}\citep{heavens}, 
does not show evidence of variability.
A possible counterpart to \igrA\ in the soft X-ray (0.3-10~keV) 
domain was identified by \citet{malizia11}. In the \swift\,/XRT field of view (FOV) 
around \igrA,\ they detected a source located at \mbox{$RA$=165.44~deg,} \mbox{$Dec$=-61.022~deg} 
(associated uncertainty $\sim$6''). By using the refined X-ray position, these authors 
also suggested that \igrA\ could be associated to the \rosat\ source 1RXH\,J110146.1-610121 
and the serendipitous \xmm\ source 2XMM\,J110147.1-61012. 

In Sect.~\ref{sec:igra} we report on the analysis of all the available X-ray observations 
performed with \swift,\ \xmm,\ \rosat,\ and \asca,\ in the direction of \igrA.\ 
We use the results of this analysis in Sect.~\ref{sec:counterparts} to search for 
possible counterparts to the source in the optical, infra-red and radio domains. 
A discussion on the nature of \igrA\ is presented in Sect.~\ref{sec:discussion}.

\section{X-ray observations of \igrA\ }
\label{sec:igra}

A summary of all X-ray observations analyzed 
in this section is provided in Table~\ref{tab:summary}. 
\begin{table}
\caption{X-ray observation log of \igrA.\ } 
\label{tab:log}
\centering
\begin{tabular}{@{}cccc@{}}
\hline\hline 
\noalign{\smallskip}
Instr &OBS ID & START TIME & Exp\\ 
        &    &  (MJD) &  (ks)   \\
\noalign{\smallskip}
\hline
\noalign{\smallskip}
 \swift &00045395002 & 55632 & 3.1\\ 
& 00045395001& 55630 & 2.1 \\ 
\noalign{\smallskip}
\xmm & 0152570101 & 52841 & 60\\ 
 & 0111210201&51753 & 11 \\ 
\noalign{\smallskip}
\rosat & RH500445A01 & 50627 & 46 \\ 
\noalign{\smallskip}
\asca &51021010 & 49733 & 40 \\ 
&51021000 & 49412 & 43\\ 
\noalign{\smallskip}
\einst & 2161 & 44463 & 11\\
 \hline
\end{tabular}
\label{tab:summary} 
\end{table}

\subsection{\swift\ }
\label{sec:swift}

\begin{figure*}
\centering
\includegraphics[width=0.945\textwidth]{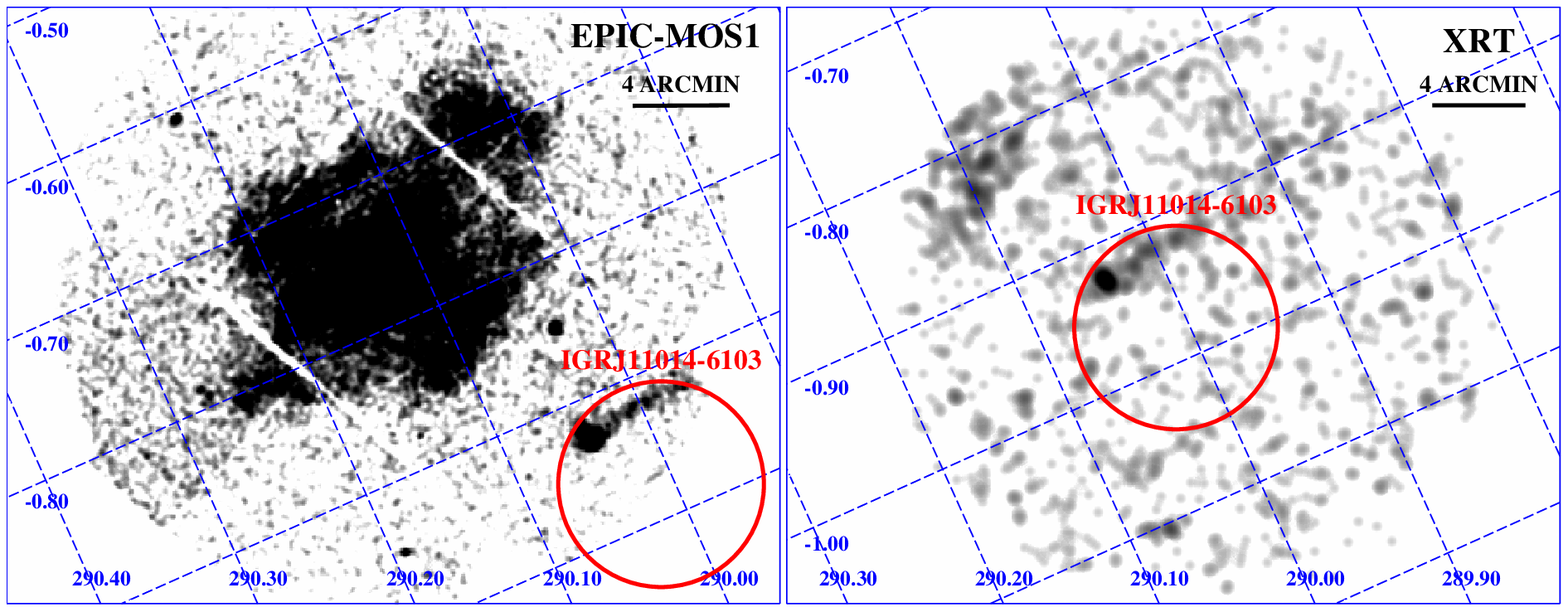}
\includegraphics[width=0.47\textwidth]{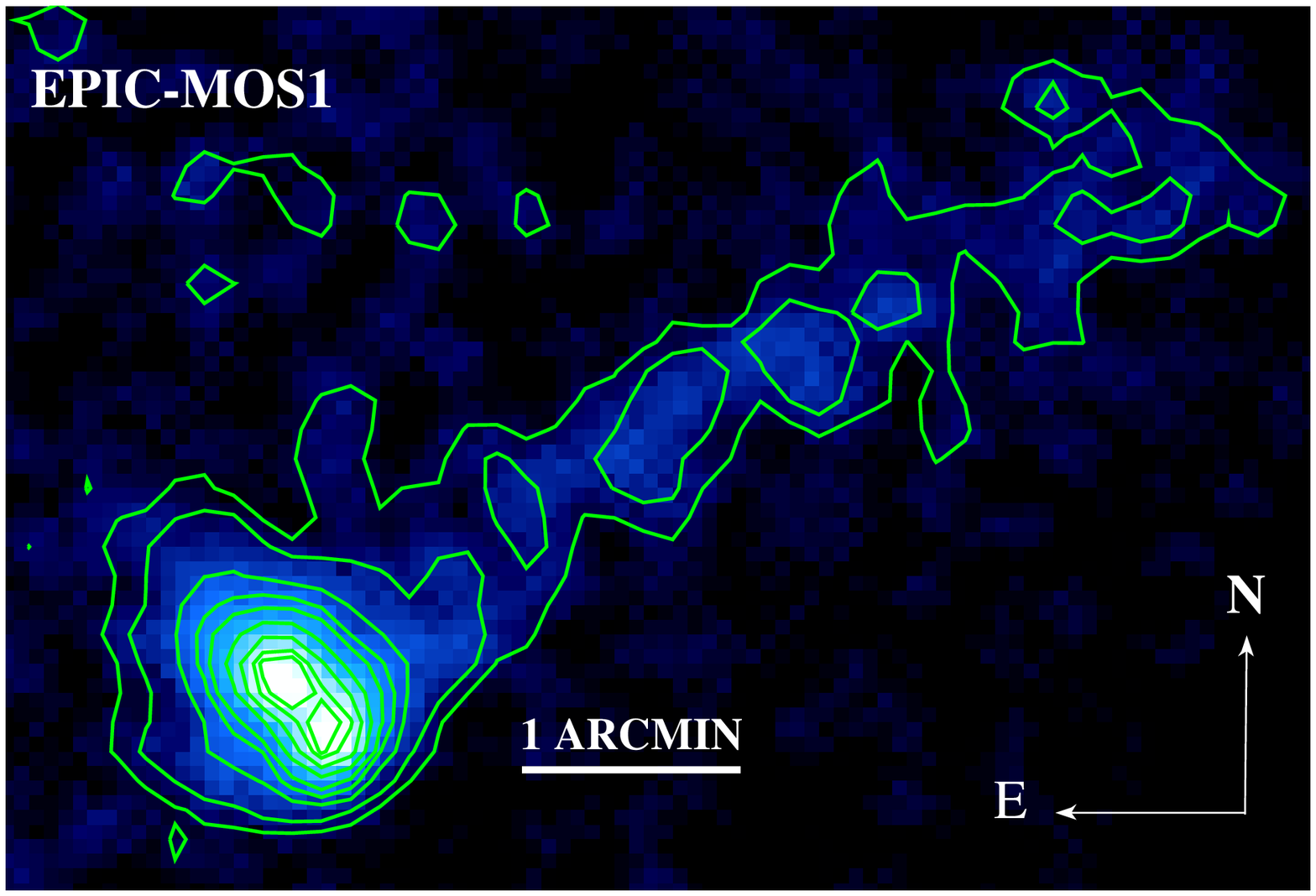}
\includegraphics[width=0.47\textwidth]{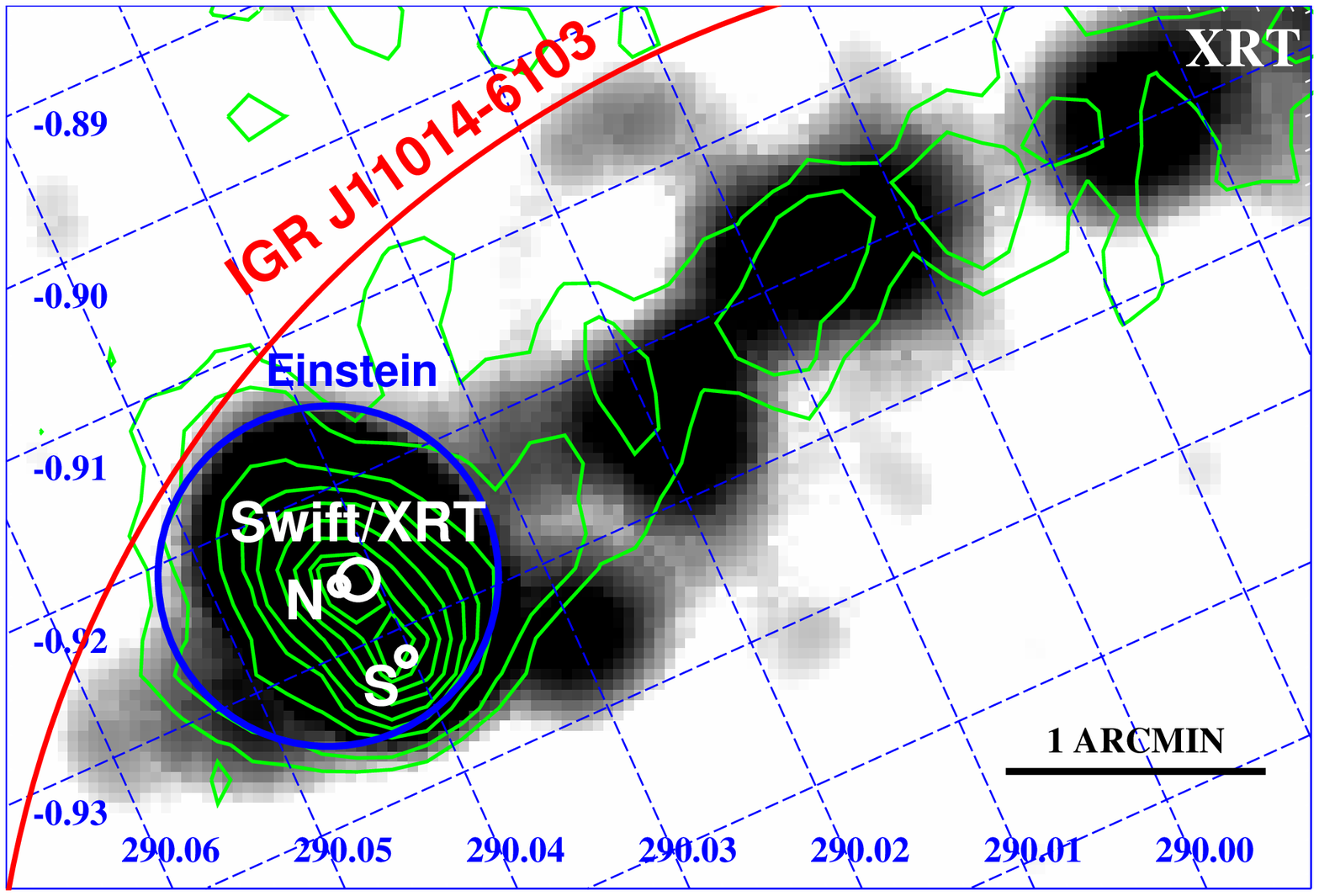}
   \caption{{\it Upper left}: FOV around \igrA\ as observed by EPIC-MOS1 
   on-board \xmm\ (0.5-12~keV). {\it Upper right}: \swift\,/XRT FOV around \igrA\ (1-9~keV). 
   {\it Bottom left}: zoom of the EPIC-MOS1 FOV. In this case we kept the colours and
   removed the coordinate grid and the error circles around the detected sources for clarity. 
   {\it Bottom right}:  zoom of the XRT FOV. In this figure we marked the source 
   detected with \swift\ (uncertainty of 4.4'' at 90\% c.l.),  
   and sources N and S detected with \xmm\ (uncertainty of 2'' at 90\% c.l.). 
   The contours determined from the \xmm\ observations are also overplotted (in green), together 
   with the error circle representing the \inte\ position of \igrA\ (in red) and the error circle of 
   the Einstein source 2E~2383 (in blue, uncertainty of 39'' at 90\% c.l., see Sect.~\ref{sec:einst}).
   The intensity scale, here and in the following images, is logarithmic and the grid is in galactic coordinates.} 
   \label{fig:xmm}
\end{figure*}

The FOV around \igrA\ was observed twice with \swift\,/XRT on 
2011 March 10 and 12 for a total exposure time of 3.1~ks and 2.1~ks, 
respectively (observations ID.~00045395001 and 00045395002). 
We analyzed the \swift\,/XRT \citep{gehrels04} data collected in 
photon counting mode (PC) by using standard procedures 
\citep{burrows05} and the latest calibration files available. 
Filtering and screening criteria were applied by using {\sc ftools}, 
and only  event grades of 0-12 were considered. 
Exposure maps were created through 
the {\sc xrtexpomap} task. 
No obvious sources were detected in the two observations separately.  
We maximized the S/N by summing-up all the available data 
(effective exposure time 5.1~ks) and extracted a single image of the XRT FOV 
around \igrA.\ In this image, we detected a faint source 
(S/N=5.5 in the 1-9~keV energy band, see Fig.~\ref{fig:xmm}) at the position 
\mbox{$RA$=165.444~deg,} \mbox{$Dec$=-61.023~deg}  
(associated uncertainty 4.4'' at 90\% c.l.; we used the {\sc xrtcentroid} tool).  
This is compatible with being the same source already identified by \citet{malizia11}, 
and is the only source detected inside the INTEGRAL error box. 
It is thus the most likely counterpart to \igrA\ in soft X-rays. 
Given the relatively short exposure time and 
the low count rate of the source, we could not extract a meaningful spectrum. Instead, we 
estimated the source count rate (0.011$\pm$0.002 counts/s, 1-9~keV) with {\sc sosta} 
(available within the ftool {\sc ximage}), 
and used this result within {\sc webpimms} in order to derive the corresponding 
X-ray flux \citep[see also][]{bozzo09}. In the conversion between count rate and flux, we adopted 
an absorbed ($N_{\rm H}$=0.7$\times$10$^{22}$~cm$^{-2}$) power law ($\Gamma$=1.6) model (see Sect.~\ref{sec:xmm}).  
The inferred X-ray flux in the 2-10~keV energy band was  
(6.3$\pm$1.2)$\times$10$^{-13}$~\ferg\  (not corrected for absorption).  
Even though a single point source is detected with the {\sc xrtcentroid} tool, 
Fig.~\ref{fig:xmm} reveals a marginal evidence for diffuse emission around the source 
and  the presence of an elongated structure ($\sim$4~arcmin) 
extending in the north-west direction from the point source. 
We discuss further these evidences in Sect.~\ref{sec:xmm}. 

\subsection{ \xmm\ }
\label{sec:xmm}

\xmm\ observed the region around \igrA\ on 28 July 2000 
(obs.~ID~0111210201, total exposure time $\sim$11~ksec) and on  
21 July 2003 (obs.~ID~0152570101, total exposure time 
$\sim 60$~ksec). In both cases, the observations were centered 
on the nearby supernova remnant (SNR) \SNR,\ and thus \igrA\ was 
off-set with respect to the instrument aim-point by about 
15~arcmin. During obs.~ID~0152570101, the EPIC-pn and EPIC-MOS cameras were operated 
in small window and full frame mode, respectively. \igrA\ was only within 
the EPIC-MOS FOV. During obs.~ID~0111210201 all the EPIC 
cameras were operated in full frame mode, and \igrA\ was observed 
with both the EPIC-pn and EPIC-MOS. In all cases, only part of the 
\inte\ error circle around \igrA\ was included in the FOV of the 
EPIC cameras (see Fig.~\ref{fig:xmm}). 

We processed the \xmm\ observation data files 
with the two pipelines {\sc epproc} and {\sc emproc} in order to produce 
EPIC-pn and EPIC-MOS cleaned event files, respectively (SAS v.10.0.1). 
To identify the high background time intervals, we followed the SAS online analysis 
thread\footnote{See also http://xmm.esac.esa.int/sas/current/documentation/threads/ 
PN\_spectrum\_thread.shtml.} and extracted the lightcurves from the full FOV of the 
EPIC cameras in the 10-12~keV energy band. We discarded from further analysis 
the time intervals in which the 10-12~keV FOV count rate was larger than 
0.4~cts/s. The resulting effective exposure times were 47.3~ks and 8.0~ks (5.7 ks) for the 
EPIC-MOS1 (EPIC-pn) cameras in the observation ID~0152570101 and 0111210201, 
respectively. All cleaned event files were barycentered with the {\em barycen} tool. 
\begin{figure}
   \centering
   \includegraphics[width=8.7cm]{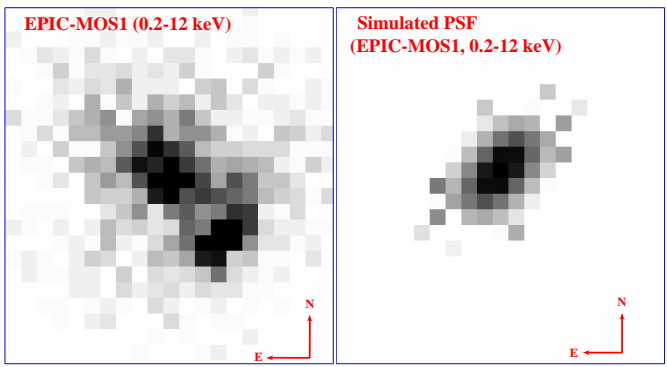}
   \caption{{\it Left}: close view of the EPIC-MOS1 FOV around 
   sources N and S (see Sect.~\ref{sec:xmm}); {\it Right}: simulated PSF of the EPIC-MOS1 
   camera at energies in the (0.2-12)~keV energy range by assuming the same off-set position as that of sources N and S. 
   The two panels have the same angular scale.
   The simulated PSF does not show the double peak profile seen in the real data.  
   We also note that the distortion due to the off-set position of the two sources 
   would occur in a different direction with respect to the separation between the two sources.}
\label{fig:psf}
\end{figure}
\begin{figure}
   \centering
   \includegraphics[width=8.7cm]{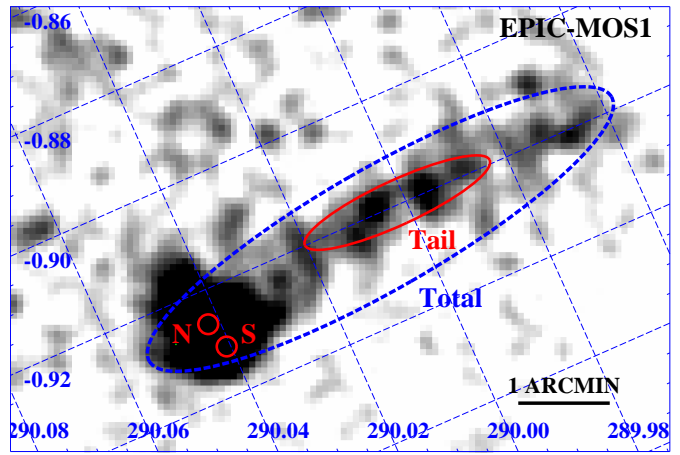}
   \caption{Zoom of the EPIC-MOS1 FOV around \igrA\ during 
   observation ID.~0152570101 (0.2-12~keV). The regions used for the 
   extraction of the spectra of sources N and S are indicated with red small circles;  
   the region used for the spectral analysis of the tail with a red ellipse. 
   We also show the extraction region used to perform the spectral analysis 
   of the entire emitting region (dashed ellipse; see Sect.~\ref{sec:xmm}).}
\label{fig:regions}
\end{figure}

A visual inspection of the EPIC-MOS image extracted from the \xmm\ 
observation with the longer exposure time (ID~0152570101, see Fig.~\ref{fig:xmm}) 
reveals that the source detected with \swift\,/XRT is indeed composed of three  
different emitting regions: two sources separated by 
$\sim 22$~arcsec (hereafter sources N and S) and a dimmer elongated structure 
($\sim$4~arcmin, hereafter, the ``tail''). In the EPIC images extracted 
from obs.~ID~0111210201 only the two brighter sources are clearly visible. 
In this observation, the tail is only marginally detectable due to the shorter 
exposure time. In order to check that sources N and S are not just produced by the distortion of the 
EPIC-MOS point spread function (PSF) at large off-set angles, we simulated with the {\em psfgen} task the 2D-PSF 
of the instrument at the location of these two sources. We used a circular region for the simulation $\sim 180$~arcsec wide
centered on source N and set the level parameter to ``ELLBETA''. 
The comparison between the simulated PSF profiles 
 and the measured intensity profiles obtained from the data
is shown in Fig.~\ref{fig:psf}. The simulated PSF did not
present a double peak in the intensity profile in a direction comparable with that 
observed from the \xmm\ data. In particular, we note that the distortion 
of the PSF caused by the off-axis position of the source on the EPIC-MOS detectors 
occurs preferably along the azimuthal direction, whereas sources N and S are 
separated in the opposite direction. 

Sources N and S were also already identified 
in the second \xmm\ catalog of serendipitous X-ray sources \citep{watson09} as 
2XMM\,J110147.1-610124 and 2XMM\,J110145.0-610140, respectively. The best determined positions are 
(\mbox{$RA$=165.4465~deg,} \mbox{$Dec$=-61.0234~deg}) for the first, and (\mbox{$RA$=165.4377~deg,} \mbox{$Dec$=-61.0279~deg}) for the latter 
\citep[the nominal positional accuracy of the EPIC cameras 
is $\sim$2'', see e.g.][and references therein]{pavan10}. 
2XMM\,J110147.1-610124 is reported as extended, and is characterized by a radial elongation of 8.1~arcsec.

In Fig.~\ref{fig:xmm} the image extracted from \swift\,/XRT is compared with that obtained from
the deeper \xmm\ observation. The contours around the extended emission from \xmm\ coincide
well with the emission in XRT.
This confirms the marginal detection of the extended emission 
noticed before in the \swift\,/XRT data (see Sect.~\ref{sec:swift}), 
and suggests that the single point source detected with this 
telescope is indeed a blend of the two brighter objects revealed by \xmm. 

\begin{figure}
   \centering
   \includegraphics[width=6.5cm,angle=-90]{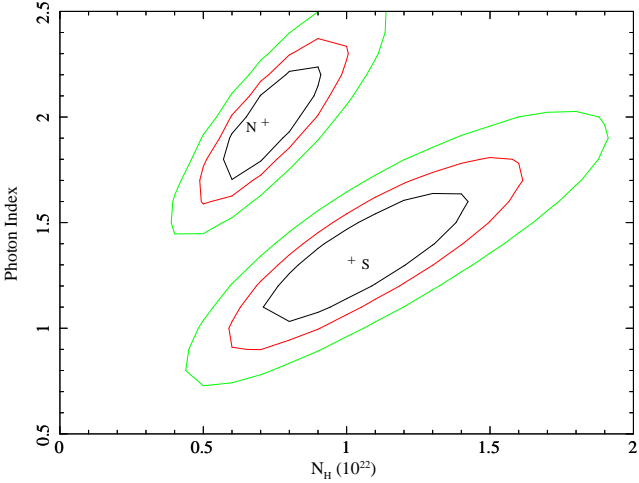}
   \caption{Contour plot of the absorption column density ($N_{\rm H}$) 
   and power law photon index ($\Gamma$) derived from the EPIC-MOS1 data for sources N and S in 
   observation ID.~0152570101. The contours correspond to 68\%, 90\% and 99\% confidence 
   levels.}
   \label{fig:contours}
\end{figure}

We also performed a spectral analysis of the X-ray emission from the two brighter sources and the 
tail extracting the corresponding spectra from the EPIC-MOS1 data in obs. ID~0152570101 
(EPIC-MOS2 data could not be used for the spectral 
analysis as in this camera the sources were located close to the gap between two different CCDs).   
The extraction regions (shown in Fig.~\ref{fig:regions}) were chosen in order to maximize the S/N 
and to minimize the reciprocal contamination.
To take into account the background contribution, we selected a circular source-free region
of radius 2~arcmin located in the same CCD as \igrA\ and at approximately the same distance from the 
central SNR. We also checked that the results presented in this paper would not be affected by 
different (reasonable) choices of the background region.
Given the relatively low number of counts, all the spectra were grouped in order to have at least 5 
photons for each energy bin and fit with an absorbed power law model 
 using C-statistics \citep[][; we used {\sc Xspec} v.12]{cash79}. We checked that similar 
results (to within the uncertainties) would have been obtained by grouping the spectra 
with 25 photons per energy bin and using $\chi^2$ statistics.  
The results of the fits  are summarized in Table~\ref{tab:xmmspec}. Here we also report the best-fit parameters
obtained from the spectral analysis of the entire region (dashed ellipse in Fig.~\ref{fig:regions}). We found marginal evidence for 
a difference between the power law photon index of sources N and S, 
and show in Fig.~\ref{fig:contours} the most relevant contour plots derived from the spectral fits.
The spectrum extracted from the tail provided only poor constraints on the properties of its 
X-ray emission. A similar analysis carried out on the EPIC-MOS 
and EPIC-pn data of the observation ID.~0111210201 did not reveal any significant variation 
of the estimated spectral parameters and fluxes from the three regions.  

\begin{figure}
   \centering
   \includegraphics[width=6.5cm,angle=-90]{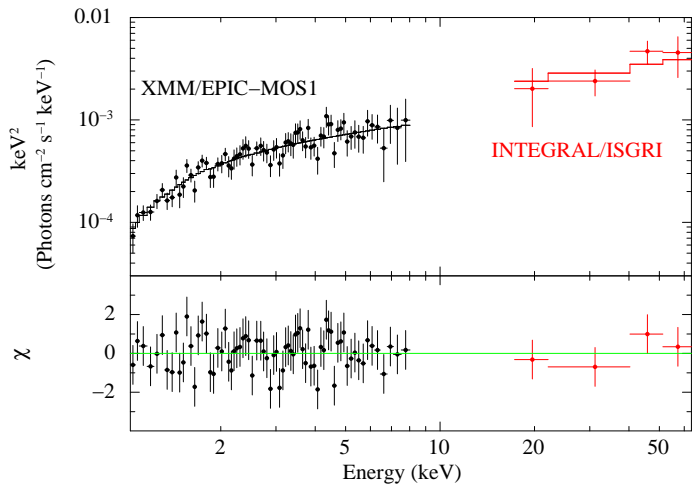}
   \caption{Unfolded \xmm\ and IBIS/ISGRI spectrum of \igrA.\ We fitted both spectra with a single absorbed 
   power law model (solid line). In the bottom panel we show the residuals from this fit.}
   \label{fig:combined}
\end{figure}

In order to compare the fluxes estimated with \xmm\ and those reported for \igrA\ in 
the 4th IBIS/ISGRI catalog (see Sect.~\ref{sec:intro}), we performed a simultaneous fit
of the MOS~1 (ID.~0152570101) spectrum from the entire source (dashed ellipse in Fig.~\ref{fig:regions})
and the IBIS/ISGRI spectrum of \igrA\ obtained 
from the {\sc Heavens} web-tool. 
\begin{table}
\caption{Results of the spectral fit parameters obtained from \xmm\ observation ID.~0152570101. 
All spectra were fit using an absorbed power law model (photon index $\Gamma$).  
The absorption column densities ($N_{\rm H}$) and fluxes are in units of 10$^{22}$~cm$^{-2}$ and 
10$^{-13}$~\ferg, respectively. We also report the estimate of the X-ray 
flux in the 0.2-2.4~keV energy band ($F_{0.2-2.4~keV}$), in order to have an easier comparison with the results 
obtained from \rosat\,/HRI observations (see Sect.~\ref{sec:rosat}). Uncertainties 
are at 90\% c.l. on the spectral parameters and 68\% c.l. on the fluxes.}
\begin{tabular}{@{}llllll@{}}
\hline
\hline
\noalign{\smallskip} 
        &  $N_{\rm H}$ & $\Gamma$ & $F_{\rm 2-10~keV}$ & $F_{\rm 0.2-2.4~keV}$ & C-stat/d.o.f.  \\
\noalign{\smallskip} 
\hline        
\noalign{\smallskip} 
N  & 0.7$\pm$0.2  & 2.0$\pm$0.3 & 3.7$^{+0.6}_{-1.1}$  & 1.2$^{+0.2}_{-0.5}$ & 68.2/68  \\
\noalign{\smallskip} 
S  & 1.0$\pm$0.4  & 1.3$\pm$0.4 & 6.2$^{+0.9}_{-2.6}$  & 0.8$^{+0.2}_{-0.4}$ &  73.3/62 \\
\noalign{\smallskip} 
Tail  &     1.3$^{+0.7}_{-0.8}$  & 2.2$^{+0.7}_{-0.8}$  &  1.9$^{+0.3}_{-1.0}$ & 0.5$^{+0.2}_{-0.3}$ & 31.1/21  \\
\noalign{\smallskip} 
Total & 0.7 $\pm$ 0.1 & 1.6 $\pm$ 0.15 & 17$^{+1.2}_{-1.4}$  & 3.9$^{+0.3}_{-0.5}$ & 68.60/85\\
\noalign{\smallskip} 
\hline
\end{tabular}
\label{tab:xmmspec}
\end{table}
The two spectra could be reasonably well fit using an absorbed power law model, with spectral parameters compatible 
with those determined previously using only the \xmm\ data (best-fit values reported in the last line of table~\ref{tab:xmmspec}). 
We obtained $N_{\rm H}$=0.6$\pm$0.2, 
$\Gamma$=1.5$\pm$0.2 ($\chi^2$/d.o.f.=0.8/74) and $C$=$1.7^{+1.1}_{ -0.6}$. Here, $C$ is 
the normalization constant introduced to take into account both the intercalibration between the EPIC-MOS1 
and IBIS/ISGRI, and any possible flux change of compact emission regions between the different observations. 
We considered EPIC-MOS1 as reference detector, therefore the constant is applied to the ISGRI spectrum.
We noticed that fixing $C$=1 would 
not significantly change the results of the fit. The estimated 2-10~keV, 20-40~keV, and 40-80~keV X-ray fluxes of the sources 
were $(1.9 \pm 0.2) \times$10$^{-12}$~\ferg , $(3.1^{+0.2}_{-0.3}) \times$10$^{-12}$~\ferg , and 
$(4.8^{+0.4}_{-0.6})\times$10$^{-12}$~\ferg , respectively. 
These are fully compatible with those reported by \citet{bird10}, i.e. 
$(3.0\pm0.8)\times$10$^{-12}$ and $(5.6\pm2)\times$10$^{-12}$~\ferg\  in the 20-40~keV and 
40-100~keV band respectively, see Sect.~\ref{sec:intro}. 
The unfolded MOS1+ISGRI spectrum 
of \igrA\ is shown in Fig.~\ref{fig:combined}. 
A simultaneous MOS1+ISGRI spectral analysis was carried out also 
using the EPIC spectra of source~S alone (without including the tail and source~N).
This source indeed seems to have an harder spectrum and higher estimated flux in the 2-10~keV band 
with respect to source~N and the tail.
No significant differences in the three spectral parameters have been measured, except for 
a marginal hint of an increase of the intercalibration constant needed in the fit.\\
\indent
We also searched in the EPIC data for indications of variability for sources N and S on time 
scales of seconds to hours, but did not spot any relevant feature. We paid particular attention 
in searching for coherent pulsations. We used the PN data for the short observation (ID.~0111210201) 
and decided to analyze sources N and S jointly, to maximize the number of source photons. 
Cutting out source S for an analysis of source N (and vice versa) would result in a too high loss 
of source photons for a detailed timing analysis, respectively. The total number of photons obtained 
for both sources together was 693. By adopting an epoch-folding technique with 16 phase bins, 
we determined an upper limit on the pulsed fraction 
for a sinusoidal signal of 56\% (at 90\% c.l.)  in the 0.002 to 6.8~Hz frequency range 
\citep[we followed the approach of][]{leahy1983}. In the long observation (ID.~0152570101) we could 
use only the data of MOS1 (1275 photons) and MOS2 (2087 photons), and explored therefore the frequency 
range 0.001--0.2~Hz: the upper limits on the pulsed fraction are 44\% and 33\% (at 90\% c.l.) respectively.

\subsection{\rosat\ }
\label{sec:rosat}

\begin{figure}
   \centering
   \includegraphics[width=8.7cm]{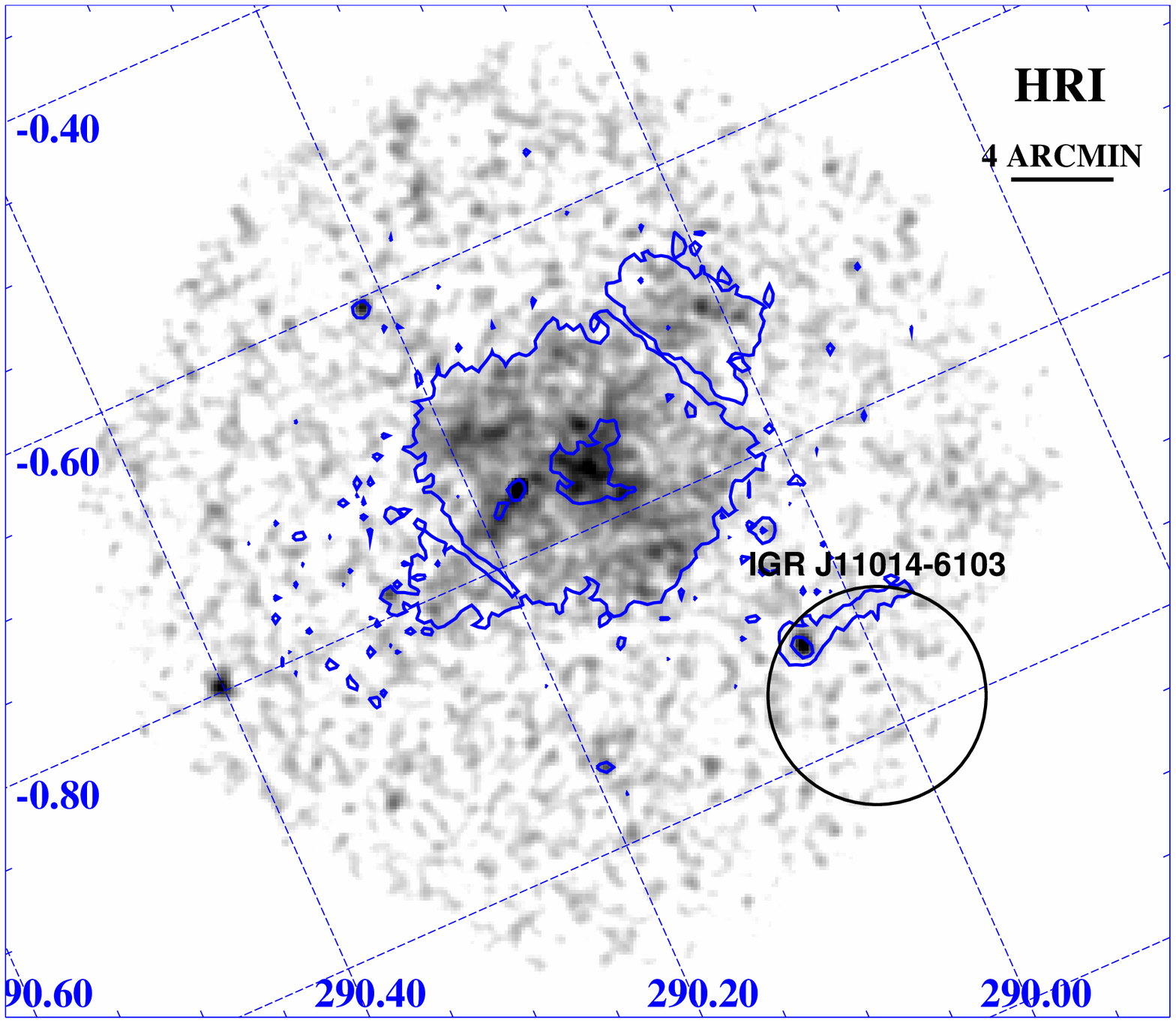}
   \includegraphics[width=8.7cm]{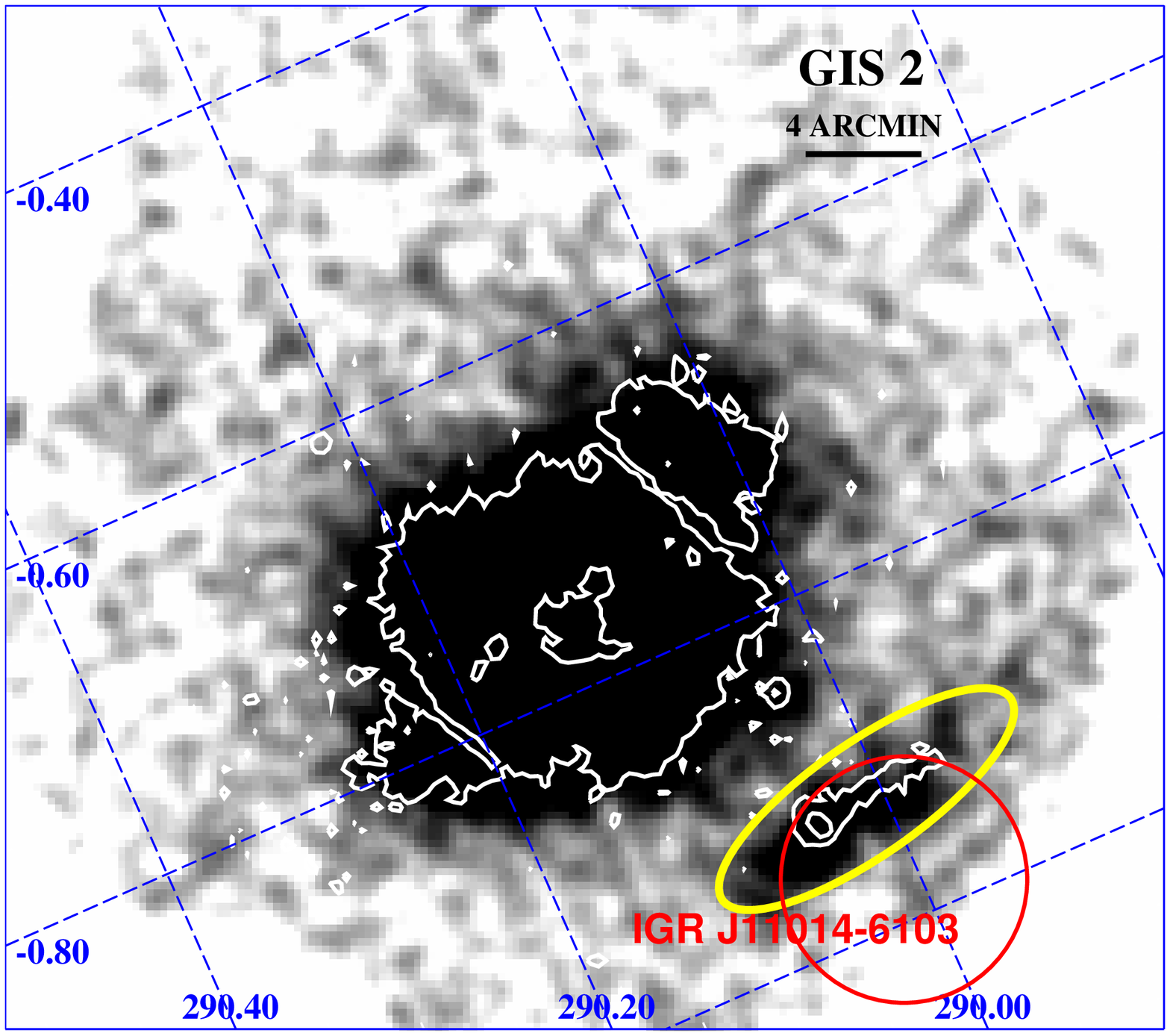}
   \caption{{\it Top}: \rosat\,/HRI FOV around \igrA\ (observation ID.~RH500445A01). 
   The image is in the 0.2-2.4 keV energy band with a resolution of 5'' (see Sect.~\ref{sec:rosat}). 
   We overplot on this image the \xmm\ contour levels (in blue) determined from the EPIC-MOS1 data 
   in observation ID.~0152570101 (see also Fig.~\ref{fig:xmm}). The black circle represents the 
   \inte\ position of \igrA.\ The bright extended emission in the center of the FOV is SNR \SNR.\ 
   {\it Bottom}: \asca\,/GIS2 FOV around \igrA\ (observation ID.~51021000).  The image is in the 0.5-10~keV energy band. 
   Extended emission is clearly visible in the bottom right side of the detector, below SNR \SNR.\ 
   We overplot on this image the \xmm\ contours (in  white), the \inte\ error circle of \igrA\ (red circle), 
   and the extraction region adopted for the GIS spectral analysis (yellow ellipse).}
\label{fig:rosat}
\end{figure}

We retrieved all available \rosat\ \citep{trumper82} data in the direction of \igrA\ from the {\sc Heasarc} 
archive. The FOV around \igrA\ was observed with the HRI \citep{pfer87} on-board \rosat\ on several occasions 
(observations ID.~RH500445A01, RH500340N00, RH500445N00, RH900619A01, and RH900619N00). Among these, 
 observation ID.~RH500445A01 was characterized by the longest available exposure time (46~ks) and 
was performed on 1997 June 28. We show in Fig.~\ref{fig:rosat} the HRI image (5'' resolution, 0.2-2.4~keV). 
Within the \inte\ error circle of \igrA\ 
a single point-like source is detected. This source  
was already classified in the HRI catalog as 1RXH\,J110146.1-610121 \citep{voges99}. 
The best determined source position is at \mbox{$RA$=165.442~deg} and \mbox{$Dec$=-61.022~deg.} The nominal positional accuracy 
of the HRI is of order 10'' \citep{predehl01} and thus the \rosat\ position is consistent with that 
of the source revealed with \swift\ and with source~N detected with \xmm.\ 
Given the limited spatial resolution, sensitivity and energy coverage 
of the HRI with respect to the EPIC cameras, it is not surprising that the closeby source~S and the tail 
are not detected in this case (source~S is also fainter than source~N in the \rosat\ energy band, 
see Table~\ref{tab:xmmspec}). We suggest that, as for the case of \swift\,/XRT, the 
\rosat\ source is indeed a blend of sources N and S.  
The estimated count rate of 1RXH\,J110146.1-610121 in the HRI energy band (0.2-2.4~keV) is 0.0038$\pm$0.0004~cts/s. 
This would convert into an X-ray flux of (9$\pm$1)$\times$10$^{-13}$~\ferg\  (2-10~keV) if an absorbed 
power law spectral model with $N_{\rm H}$=0.7$\times$10$^{22}$~cm$^{-2}$ 
and $\Gamma$=1.6 is used (see Sect.~\ref{sec:xmm}). The flux of the \rosat\ source is thus 
qualitatively in agreement with that of sources N and S estimated from \xmm\ data. 

We noted that the FOV around \igrA\ was observed also with the PSPC on-board \rosat\ on two occasions, 
on 1993 July 23 (observation ID.~RP500307N00) and on 1994 July 5 (observation ID.~RP500307A01). 
However, in these two cases the relative low exposure time ($\sim$2~ks) and the off-set position of the source 
with respect to the instrument aim-point, did not result in a clear detection of the source. 
We thus do not discuss these data in more detail. 

\subsection{\asca\ }
\label{sec:asca}

In the {\sc Heasarc} archive we found two observations carried out with the 
GIS telescope \citep{ohashi96} on-board \asca\ \citep{tanaka94} which included the region 
around \igrA\ in their FOV.  The first of these observations, 
ID.~51021000, was performed on 1994 March 1 with a total exposure 
time of 43.4~ks. The second observation, ID.~51021010, lasted for  40.5~ks and was carried out 
about 10 months later, on 1995 January 16. 
Both these observations were aimed at the SNR \SNR\ and did not   
cover entirely the \inte\ error circle around the position of \igrA.\  
However, the extended emitting region already identified with \xmm\ is clearly detected in  
the GIS images (see Fig.~\ref{fig:rosat}). 
In comparison with the HRI camera on-board \rosat,\ the GIS has a broader energy coverage 
(0.5-10~keV) and a larger sensitivity, but a significantly worse spatial resolution 
\citep[the GIS PSF is $\sim$1~arcmin and strongly dependent on energy,][]{tanaka94}. 
It is thus not surprising that the GIS is not able to distinguish the two sources (N and S)  
detected with \xmm\ and \rosat,\ but can reveal the presence of the diffuse emission around them. 

We analyzed GIS2 and GIS3 data following standard procedures and the latest available 
calibration files \citep[see e.g.][and references therein]{pavan10}. 
Source spectra in the two observations were extracted using the elliptical region shown 
in Fig.~\ref{fig:rosat}. Given the presence of the nearby SNR~\SNR,\ we extracted 
the background spectra from different regions located at the same radial distances of \igrA\ 
from \SNR.\ Even though the S/N of the data was relatively poor, the GIS spectra 
could be reasonably well fit using an absorbed power law model. From the GIS2 data in observation 
51021000, we measured $\Gamma$=1.6$\pm$0.3 and estimated a 2-10~keV X-ray flux of 
(2.1$^{+0.2}_{-0.4}$)$\times$10$^{-12}$~\ferg\  ($\chi^2_{\rm red}$/d.o.f.=0.7/21). 
These values are compatible (to within the errors) with those inferred from the \xmm\ data.  
From observation~51021010, we obtained $\Gamma$=1.2$\pm$0.5 and 
$F_{\rm 2-10~keV}$=(2.1$^{+0.3}_{-1.2}$)$\times$10$^{-12}$~\ferg\  
($\chi^2_{\rm red}$/d.o.f.=0.9/16). In both cases we fixed the absorption column density 
to the best value measured from \xmm.\ Compatible results (to within the uncertainties) 
were obtained from the GIS3 data. 

\subsection{\einst\ }
\label{sec:einst}

The region around \igrA\ was observed also with the \einst/IPC 
detector~\citep[0.2-4.5~keV, ][]{Giacconi:1979lr} on 1980 August 12 with an exposure time $\sim 11$ks.
A visual inspection of the \einst/IPC image revealed a source (2E~2383) spatially coincident 
with sources~N and S and a marginal evidence for an extended tail similar to that observed in \xmm.
The best position of 2E~2383 is reported in the 2E~catalog at \mbox{$RA=11^{\mathrm{h}}:01^{\mathrm{m}}:47.5^{\mathrm{s}}$,} 
\mbox{$Dec$=-61$^\circ$:01':22''}
\citep[related uncertainty $39\arcsec$, see Fig.\ref{fig:xmm}; ][]{harris94}.
The source count rate was 0.011$\pm$0.002~cts/s, which correspond to $(9\pm2)\times 10^{-13}$~\ferg\ 
(assuming a column density $N_H=0.7 \times 10^{22}$~cm$^{-2}$ and a power law spectrum with slope $\Gamma=1.6$, 
~see Sect.~\ref{sec:xmm}).
This flux is compatible with the one measured with \xmm.

\section{Counterparts to \igrA\ }
\label{sec:counterparts}

In this section we take advantage of the improved positions for the 
three emitting regions comprised in the \inte\ error circle around \igrA\ 
to search for the corresponding counterparts in the optical, infra-red, 
and radio domain. In Sect.~\ref{sec:gamma}, we also report on 
a possible $\gamma$-ray detection.

\subsection{Optical/IR data}
\label{sec:opt-ir}

\begin{figure}
   \centering
   \includegraphics[width=8.7cm]{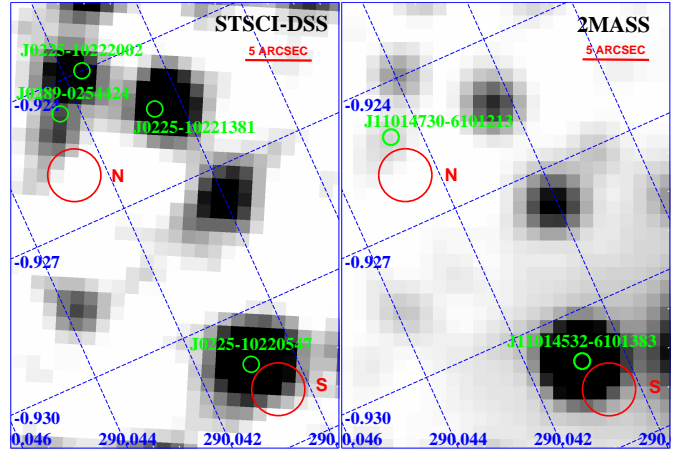}
   \caption{Search for optical and infra-red counterparts to the 
   sources N and S. {\it Left}: STCI-DSS image. We overplot the position 
   of sources N and S (radius equal to the position accuracy, 2'') 
   as determined from the EPIC-MOS1 observation, and the most likely optical counterpart 
   from the USNO-A2.0 and USNO-B1.0 catalogs (green circles, with radius 0.6''). 
   {\it Right}: 2MASS image. The position of  
   sources N and S is indicated as above. We marked the location of 
   the potential infra-red counterparts retrieved from the 2MASS catalog 
   (green circles, with radius 0.6'').}
   \label{fig:otticoNIR}
\end{figure}

We used publicly available optical and near infra-red images from
 STSCI-DSS,  and 2MASS surveys.
Potential counterparts to sources~N and S are shown in 
Fig.~\ref{fig:otticoNIR}. For source N, we did not find any obvious 
counterpart in the optical domain. The closest catalogued USNO-A2.0 sources 
are located more than $\sim$7'' away from the best determined EPIC-MOS1 position of the 
source. The nearest classified object in the USNO-B1.0 catalog is 
J0289-0254424 (R1=10.89, B2=15.97), located at \mbox{$RA$=165.4471~deg,} 
\mbox{$Dec$=-61.0221~deg.} In the infra-red domain, the closest catalogued source is 
the 2MASS object J11014730-6101213 located at \mbox{$RA$=165.447122~deg,} \mbox{$Dec$=-61.022602~deg.} 
This object is reported in the catalog with J=15.95, H=15.54, K=15.46. 
For source~S, the closest catalogued source in the optical domain (USNO-A2.0 catalog) 
is J0225-10220547. This is located at \mbox{$RA$=165.438870~deg,} 
\mbox{$Dec$=-61.027375~deg} and is characterized by B=16.5 and R=13.0. 
In the infra-red domain, the closest object from the 2MASS catalog 
to the position of source S is J11014532-6101383, located 
at \mbox{$RA$=165.438860~deg,} \mbox{$Dec$=-61.027306~deg} and characterized by J=11.16, H=10.19, K=9.89. 

\subsection{Radio data}
\label{sec:radio}

\begin{figure}
   \centering
   \includegraphics[width=0.5 \textwidth]{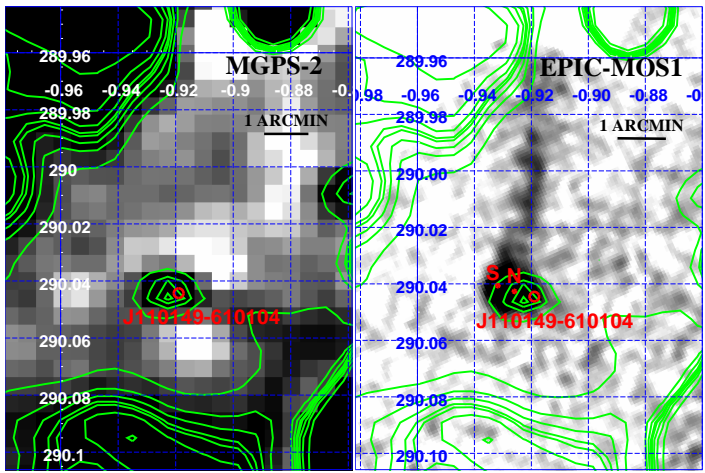}
   \caption{{\it Left}: radio image (and contours) at 843~MHz around the position of \igrA\ retrieved from 
   the MGPS-2 archive. We mark in red the position of the source MGPS2\,J110149-610104. 
   {\it Right:} EPIC-MOS1 image from observation ID.~0152570101. We overplot on the image the radio  
    contours derived from the MGPS-2 data in green and marked in red the position of the radio source 
    J110149-610104 and the two \xmm\ sources N and S.}
\label{fig:mgps1}
\end{figure}

To search for radio counterparts to \igrA,\ we used data 
from the MGPS-2 archive. The MGPS-2 is a high-resolution 
and large-scale survey of the galactic plane carried out with the Molonglo 
Observatory Synthesis Telescope (MOST) at a frequency of 843~MHz 
\citep{Murphy:2007lr}. The MGPS-2 image covering the field around 
\igrA\ is shown in Fig.~\ref{fig:mgps1}. 
We noticed from this figure the presence of a relatively 
bright radio source, J110149-610104, positionally consistent 
with sources N and S.  
The radio source is reported in the MGPS catalog as ``compact'', being  
its dimensions comparable to the convoluted beam of the survey.
The best determined position of J110149-610104 is \mbox{$RA$=(165.4559$\pm$0.0017)~deg,} 
\mbox{$Dec$=(-61.0178$\pm$0.0017)~deg.}  
The associated flux at 843~MHz is 24.2$\pm$4.8~mJy 
(corresponding to (2.04$\pm$0.40) $\times 10^{-16}$~\ferg ). 

\subsection{$\gamma$-ray data}
\label{sec:gamma}

We also noticed the presence of an \egret\ source, 3EG~J1102-6103, which might be related 
with \igrA.\ In the latest available \egret\ catalog, the best determined position of 
3EG~J1102-6103 is at \mbox{$RA$=165.60~deg,} \mbox{$Dec$=-61.05~deg,} with an associated uncertainty of 
0.6~deg. The error circle around 3EG~J1102-6103 would thus be compatible 
with the position of \igrA,\ as determined before. At present, however, 
we are unable to confirm this association due to the presence of other possible 
counterparts located within the \egret\ error circle of 3EG~J1102-6103. Among these,  
there are the SNR \SNR\ and the young energetic pulsar PSR~J1105-6107 \citep{Kaspi:1997lr}. 

In the \fermi/LAT catalog we did not find any source positionally coincident with \igrA. 
We did not attempt an estimate of an upper limit on the source flux from the \fermi\ data.
The latter calculation would indeed depend crucially on the properties of the still relatively poorly known high energy Galactic background
in such a highly crowded region.

\section{Discussion and conclusions}
\label{sec:discussion}

In this paper we analyzed all publicly available X-ray observations carried out 
in direction of the unidentified source \igrA\ with \swift, \xmm, \rosat, \asca\ and \einst.
The entire \inte\ error circle around \igrA\ has been covered in the soft
(1-10~keV) X-ray energy band with \swift\ and \rosat. 
The analysis of these data led to the detection of a single X-ray source 
 inside the error circle given by \inte\ \citep[see also][]{malizia11}.
However, the deepest observations of the region around 
\igrA, performed with the EPIC cameras on-board \xmm, revealed
that the source detected with \swift/XRT and \rosat\ is likely a blend of two distinct objects:
a point-like source and an 8.1~arcsec extended object.
The same observations revealed also the presence of a cometary-like tail ($\sim$4~arcmin) 
extending from the two sources, and comprised within the \inte\ error circle around \igrA.
The X-ray spectra of the three objects displayed only marginally significant differences 
(see Table~\ref{tab:xmmspec} and  Fig.~\ref{fig:contours}).\\
The \rosat\,/HRI and \asca\,/GIS observations of the same region 
were unable to disentangle the three emitting components. 
This can be ascribed to the narrower energy range coverage of the former 
instrument and the lower spatial resolution of the telescopes on board \asca\ and 
\rosat\ with respect to the EPIC cameras (see fluxes in the 0.2-2.4~keV band in 
Table~\ref{tab:xmmspec} and Sect.~\ref{sec:rosat},~\ref{sec:asca}). 
In all cases, we showed that the X-ray emitting 
properties of the entire region comprised in the \inte\ error circle of \igrA\ 
were in agreement among all the available instruments. 
In Sect.~\ref{sec:xmm}, we also showed that the joint \xmm\/+\inte\ 
spectrum could be reasonably well described using a single absorbed power law 
model, with a normalization constant between the two instruments $\simeq 1$.
All these results suggest that \igrA\ is a persistent X-ray emitter that displayed in the 
past $\sim$ 31 years only marginal variations (if any) in the 0.3-100~keV energy band. 

The improved X-ray positions of sources N and S permitted also to search for 
their possible optical and infra-red counterparts. The closest catalogued 2MASS, USNO-A2.0, and 
USNO~B1.0 objects to sources~N and S lie just outside the  
90\% c.l. error circle determined with \xmm; further observations with the 
ACIS telescope on-board \chan\ are therefore needed to reduce the likelihood of a chance coincidence.
We found a possible radio counterpart close to sources N and S. The radio contours 
derived from the publicly available MGPS-2 data match reasonably well with those of 
the X-ray emission detected with \xmm\ around sources N 
and S. The presence of an \egret\ source, 3EG~J1102-6103, positionally
coincident with \igrA\ was also reported.\\

Being \igrA\ located only 11~arcmin apart from the nearby SNR~\SNR,\ 
 we first considered the possibility that the X-ray emission coming from 
 the unidentified \inte\ source was related to the SNR. However in none of
the investigated wavelengths (X-ray, radio, optical, NIR)
we could clearly identify the presence of emitting 
structures located between \SNR\ and \igrA\ which could have favored 
such an association \citep[see also][]{rosado1996}.

As an alternative and perhaps more viable hypothesis, we considered that the morphology and 
emission properties of the 4~arcmin extended X-ray tail in \igrA\
might resemble the elongated features observed in the case of 
PSR~B2224+65 \citep[``the Guitar'' nebula,][]{hui2007x}
and PSR~J0357+3205 \citep{deluca2011}.
In both cases a pulsar (PSR) has been firmly detected
at one end of the elongated structures, and
the cometary-like tails have been tentatively explained under the
 ``bow-shock pulsar wind nebulae'' (bsPWN) scenario. 
These elongated cometary-like objects form when a  
pulsar escapes from the associated supernova remnant and moves with a very 
high velocity through the interstellar medium \citep[typically hundreds of km/s, i.e. 
orders of magnitude higher than the velocities of the 
cold and warm components of the interstellar medium, see e.g.]
[and references therein]{roberts05}. Close to the pulsar, the relativistic 
wind of the compact object usually gives rise to a sub-luminous cavity 
in X-rays surrounded by a termination shock. In this region, particles 
are thermalized and accelerated, thus producing a conspicuous X-ray and 
radio emission (through synchrotron processes due to the interaction 
with the local magnetic field). As the pulsar is moving, a bow shock is 
formed in front of it and the flow of material advects the emitting 
particles back along the direction of motion, leading to the formation 
of an elongated cometary structure 
\citep[see e.g.][for a detailed discussion]{gaensler2006}. 
In a few cases, deep X-ray observations with the ACIS telescope on-board 
\chan\ permitted to clearly disentangle the different contributions to the 
total X-ray and radio emission produced by the structures around a bsPWN 
\citep[see e.g. the case of the PWN G359.23-0.82;][]{gaensler2004}.

If the structures detected in \igrA\ are indeed originating from the pulsar wind, the X-ray tail
can be interpreted as being due to relic electrons spread along the direction of motion of the PSR.
Source~N, extended in X-rays and closer to the peak of the radio emission,
would represent, according to this interpretation, the compact PWN 
(with possible contribution coming also from the bow-shock).
The detection of a radio source located close to source~N would support  
 this scenario, as radio synchrotron 
emission is expected from particles accelerated at the termination shock; 
see the cases of, e.g. PSRs B1853+01 \citep{frail1996, petre2002}, 
B1957+20 \citep{stappers2003}, B1757-24 \citep{frail1991, kaspi2001}.

While the association between the extended source~N ad the ``tail'' seems morphologically plausible,
the association of source~S to this region is less straightforward.

In case source~S is not physically related to the X-ray tail,
it might just by chance be located along the line of sight to the extended structure.
The pulsar responsible for the formation of the compact and relic PWNe,
might, in this case, be unresolved and comprised in the emission detected from source~N.

In case, instead, that source~S, the point-like source, is physically connected to the PWN,
it would be reasonable to associate 
it to the PSR generating the X-ray structures.
Under this scenario the misalignment between the direction of the
4~arcmin tail and the axis between sources~N and S
would  imply a significant tilt between the proper motion of the PSR and
the position of the compact PWN.
A similar misalignment, however, was already observed in e.g. PSR~B2224+65,
 and ascribed to the strong interaction between the fast-moving pulsar and its  
dense surrounding environment \citep[][]{hui2007x}.
Indeed, as for the case of PSR~B2224+65, 
also the environment around \SNR\ and \igrA\ was found to be
 a complex and dense region, characterized by the presence of several 
background and foreground molecular clouds \citep[][and references therein]{filipovic2005}.

Following the joint XMM+ISGRI spectral analysis (see Sect.~\ref{sec:xmm}),
we conservatively assume that both sources~N and S contribute to the ISGRI emission
(the tail component has indeed a much lower flux).
Spectral properties (hardness and flux stability) similar to those observed in 
the present case (see Fig.~\ref{fig:combined}) have 
 already been measured in, e.g. AX~J1838-0655. 
This source is a young ($\tau_c \equiv P/2\dot{P} = 23$~kyr) pulsar, 
surrounded by a PWN \citep{gotthelf2008}. 
The best-fit power law slope in the 2-100~keV was measured at $\Gamma=1.5+/-0.2$ \citep{maliziaAXJ1838.0-0655}.
We note however that also the highly-magnetized neutron stars 
\citep[``magnetars'', for a review see e.g. ][]{mereghetti2008}
 share similar spectral properties. 
SGR~1806-20, for example, showed a power law slope $\Gamma=1.6-2.0$ in the range 1-100~keV 
\citep{esposito2007}.
As the 1-10~keV spectra of magnetars are usually (phenomenologically) 
modelled with a combined black body+power law (BB+PL) model \citep{mereghetti2008},
the lack of the BB component in the \xmm\ spectrum might 
be an issue for this interpretation.
We therefore estimated for source~S (i.e. the component showing the higher 2-10~keV flux) 
the contribution of an eventual 
black body spectral component. We used a temperature 
$kT=0.5$, in agreement with the one measured for SGR~1806-20 \citep{esposito2007}
and for magnetars in general. 
Due to the relative low S/N, we could only obtain a 
poorly constrained upper limit on the ratio 
between the fluxes of the BB and PL components $F_{\rm BB} / F_{\rm PL} < 0.1$.
This would be compatible with the value measured in the case of 
SGR~1806-20 \citep[$F_{\rm BB} / F_{\rm PL} =0.03$, ][]{esposito2007}.
The present data therefore do not argue against a magnetar hypothesis.
We note also that a PWN associated to a magnetar has already been observed, 
e.g. in the case of AXP~1E1547.0-5408 \citep{vink2009}.\\

Further observations with an X-ray telescope characterized by a finer spatial 
resolution (i.e. the ACIS on-board \chan) and 
increased spectral and timing statistics
 are required to solve the issues above and firmly establish the real nature of \igrA. 
In case these observations will 
confirm that \igrA\ is a newly discovered PWN generated by a high-velocity PSR, 
this would be the first detection with \inte\ 
of one of these systems (to the best of our knowledge).

\begin{acknowledgements} 
This work made use of data obtained from the High Energy 
Astrophysics Science Archive Research
Center ({\sc Heasarc}), provided by NASA’s Goddard Space Flight Center. 
We thank an anonymous referee for useful comments.
\end{acknowledgements}

\bibliographystyle{aa}
\bibliography{pavan-igrJ11}

\begin{thebibliography}{40}
\expandafter\ifx\csname natexlab\endcsname\relax\def\natexlab#1{#1}\fi

\bibitem[{{Bird} {et~al.}(2010){Bird}, {Bazzano}, {Bassani}, {Capitanio},
  {Fiocchi}, {Hill}, {Malizia}, {McBride}, {Scaringi}, {Sguera}, {Stephen},
  {Ubertini}, {Dean}, {Lebrun}, {Terrier}, {Renaud}, {Mattana}, {G{\"o}tz},
  {Rodriguez}, {Belanger}, {Walter}, \& {Winkler}}]{bird10}
{Bird}, A.~J., {Bazzano}, A., {Bassani}, L., {et~al.} 2010, \apjs, 186, 1

\bibitem[{{Bozzo} {et~al.}(2009){Bozzo}, {Giunta}, {Stella}, {Falanga},
  {Israel}, \& {Campana}}]{bozzo09}
{Bozzo}, E., {Giunta}, A., {Stella}, L., {et~al.} 2009, \aap, 502, 21

\bibitem[{{Burrows} {et~al.}(2005){Burrows}, {Hill}, {Nousek}, {Kennea},
  {Wells}, {Osborne}, {Abbey}, {Beardmore}, {Mukerjee}, {Short}, {Chincarini},
  {Campana}, {Citterio}, {Moretti}, {Pagani}, {Tagliaferri}, {Giommi},
  {Capalbi}, {Tamburelli}, {Angelini}, {Cusumano}, {Br{\"a}uninger}, {Burkert},
  \& {Hartner}}]{burrows05}
{Burrows}, D.~N., {Hill}, J.~E., {Nousek}, J.~A., {et~al.} 2005, \ssr, 120, 165

\bibitem[{{Cash}(1979)}]{cash79}
{Cash}, W. 1979, \apj, 228, 939

\bibitem[{{Chaty} {et~al.}(2010){Chaty}, {Zurita Heras}, \&
  {Bodaghee}}]{chaty10}
{Chaty}, S., {Zurita Heras}, J.~A., \& {Bodaghee}, A. 2010,
  ArXiv:astro-ph/1012.2318

\bibitem[{De~Luca {et~al.}(2011)De~Luca, Marelli, Mignani, Caraveo, Hummel,
  Collins, Shearer, Parkinson, Belfiore, \& Bignami}]{deluca2011}
De~Luca, A., Marelli, M., Mignani, R., {et~al.} 2011, Arxiv preprint
  arXiv:1102.3278, The Astrophysical Journal, 35

\bibitem[{Esposito {et~al.}(2007)Esposito, Mereghetti, Tiengo, Zane, Turolla,
  G\"otz, Rea, Kawai, Ueno, Israel, Stella, \& Feroci}]{esposito2007}
Esposito, P., Mereghetti, S., Tiengo, A., {et~al.} 2007, A\&A, 476, 321

\bibitem[{Filipovic {et~al.}(2005)Filipovic, Payne, \& Jones}]{filipovic2005}
Filipovic, M., Payne, J., \& Jones, P. 2005, Serbian Astronomical Journal, 170,
  47

\bibitem[{{Frail} {et~al.}(1996){Frail}, {Giacani}, {Goss}, \&
  {Dubner}}]{frail1996}
{Frail}, D.~A., {Giacani}, E.~B., {Goss}, W.~M., \& {Dubner}, G. 1996, \apjl,
  464, L165

\bibitem[{{Frail} \& {Kulkarni}(1991)}]{frail1991}
{Frail}, D.~A. \& {Kulkarni}, S.~R. 1991, \nat, 352, 785

\bibitem[{Gaensler \& Slane(2006)}]{gaensler2006}
Gaensler, B. \& Slane, P. 2006, ARA\&A, 44, 17

\bibitem[{Gaensler {et~al.}(2004)Gaensler, Swaluw, Camilo, Kaspi, Baganoff,
  Yusef-Zadeh, \& Manchester}]{gaensler2004}
Gaensler, B., Swaluw, E., Camilo, F., {et~al.} 2004, ApJ, 616, 383

\bibitem[{{Gehrels} {et~al.}(2004){Gehrels}, {Chincarini}, {Giommi}, {Mason},
  {Nousek}, {Wells}, {White}, {Barthelmy}, {Burrows}, {Cominsky}, {Hurley},
  {Marshall}, {M{\'e}sz{\'a}ros}, {Roming}, {Angelini}, {Barbier}, {Belloni},
  {Campana}, {Caraveo}, {Chester}, {Citterio}, {Cline}, {Cropper}, {Cummings},
  {Dean}, {Feigelson}, {Fenimore}, {Frail}, {Fruchter}, {Garmire}, {Gendreau},
  {Ghisellini}, {Greiner}, {Hill}, {Hunsberger}, {Krimm}, {Kulkarni}, {Kumar},
  {Lebrun}, {Lloyd-Ronning}, {Markwardt}, {Mattson}, {Mushotzky}, {Norris},
  {Osborne}, {Paczynski}, {Palmer}, {Park}, {Parsons}, {Paul}, {Rees},
  {Reynolds}, {Rhoads}, {Sasseen}, {Schaefer}, {Short}, {Smale}, {Smith},
  {Stella}, {Tagliaferri}, {Takahashi}, {Tashiro}, {Townsley}, {Tueller},
  {Turner}, {Vietri}, {Voges}, {Ward}, {Willingale}, {Zerbi}, \&
  {Zhang}}]{gehrels04}
{Gehrels}, N., {Chincarini}, G., {Giommi}, P., {et~al.} 2004, ApJ, 611, 1005

\bibitem[{Giacconi {et~al.}(1979)Giacconi, Branduardi, Briel, Epstein,
  Fabricant, Feigelson, Forman, Gorenstein, Grindlay, Gursky, Harnden, Henry,
  Jones, Kellogg, Koch, Murray, Schreier, Seward, Tananbaum, Topka,
  Van~Speybroeck, Holt, Becker, Boldt, Serlemitsos, Clark, Canizares, Markert,
  Novick, Helfand, \& Long}]{Giacconi:1979lr}
Giacconi, R., Branduardi, G., Briel, U., {et~al.} 1979, ApJ, 230, 540

\bibitem[{Gotthelf \& Halpern(2008)}]{gotthelf2008}
Gotthelf, E.~V. \& Halpern, J.~P. 2008, ApJ, 681, 515

\bibitem[{Harris {et~al.}(1994)Harris, Forman, Gioa, Hale, Harnden, Jones,
  Karakashian, Maccacaro, McSweeney, Primnini, Schwarz, Tananbaum, \&
  Thurman}]{harris94}
Harris, D., Forman, W., Gioa, I., {et~al.} 1994, EINSTEIN Observatory catalog
  of IPC X-ray sources, SAO HEAD CD-ROM Series I (Einstein)

\bibitem[{Hui \& Becker(2007)}]{hui2007x}
Hui, C. \& Becker, W. 2007, A\&A, 467, 1209

\bibitem[{Kaspi {et~al.}(1997)Kaspi, Bailes, Manchester, Stappers, Sandhu,
  Navarro, \& D'Amico}]{Kaspi:1997lr}
Kaspi, V.~M., Bailes, M., Manchester, R.~N., {et~al.} 1997, ApJ, 485, 820

\bibitem[{{Kaspi} {et~al.}(2001){Kaspi}, {Gotthelf}, {Gaensler}, \&
  {Lyutikov}}]{kaspi2001}
{Kaspi}, V.~M., {Gotthelf}, E.~V., {Gaensler}, B.~M., \& {Lyutikov}, M. 2001,
  \apjl, 562, L163

\bibitem[{{Leahy} {et~al.}(1983){Leahy}, {Darbro}, {Elsner}, {Weisskopf},
  {Kahn}, {Sutherland}, \& {Grindlay}}]{leahy1983}
{Leahy}, D.~A., {Darbro}, W., {Elsner}, R.~F., {et~al.} 1983, \apj, 266, 160

\bibitem[{Malizia {et~al.}(2005)Malizia, Bassani, Stephen, Bazzano, Ubertini,
  Bird, Dean, Sguera, Renaud, Walter, \& Gianotti}]{maliziaAXJ1838.0-0655}
Malizia, A., Bassani, L., Stephen, J.~B., {et~al.} 2005, ApJ, 630, L157

\bibitem[{{Malizia} {et~al.}(2011){Malizia}, {Landi}, {Bassani}, {Bazzano},
  {Bird}, {Gehrles}, \& {Kennea}}]{malizia11}
{Malizia}, A., {Landi}, R., {Bassani}, L., {et~al.} 2011, The Astronomer's
  Telegram, 3290, 1

\bibitem[{{Mereghetti}(2008)}]{mereghetti2008}
{Mereghetti}, S. 2008, A\&Ar, 15, 225

\bibitem[{Murphy {et~al.}(2007)Murphy, Mauch, Green, Hunstead, Piestrzynski,
  Kels, \& Sztajer}]{Murphy:2007lr}
Murphy, T., Mauch, T., Green, A., {et~al.} 2007, MNRAS, 382, 382

\bibitem[{{Ohashi} {et~al.}(1996){Ohashi}, {Ebisawa}, {Fukazawa}, {Hiyoshi},
  {Horii}, {Ikebe}, {Ikeda}, {Inoue}, {Ishida}, {Ishisaki}, {Ishizuka},
  {Kamijo}, {Kaneda}, {Kohmura}, {Makishima}, {Mihara}, {Tashiro}, {Murakami},
  {Shoumura}, {Tanaka}, {Ueda}, {Taguchi}, {Tsuru}, \& {Takeshima}}]{ohashi96}
{Ohashi}, T., {Ebisawa}, K., {Fukazawa}, Y., {et~al.} 1996, PASJ, 48, 157

\bibitem[{{Pavan} {et~al.}(2011){Pavan}, {Bozzo}, {Ferrigno}, {Ricci},
  {Manousakis}, {Walter}, \& {Stella}}]{pavan10}
{Pavan}, L., {Bozzo}, E., {Ferrigno}, C., {et~al.} 2011, \aap, 526, A122

\bibitem[{{Petre} {et~al.}(2002){Petre}, {Kuntz}, \& {Shelton}}]{petre2002}
{Petre}, R., {Kuntz}, K.~D., \& {Shelton}, R.~L. 2002, \apj, 579, 404

\bibitem[{{Pfeffermann} {et~al.}(1987){Pfeffermann}, {Briel}, {Hippmann},
  {Kettenring}, {Metzner}, {Predehl}, {Reger}, {Stephan}, {Zombeck}, \&
  {Chappell}}]{pfer87}
{Pfeffermann}, E., {Briel}, U.~G., {Hippmann}, H., {et~al.} 1987, in Presented
  at the Society of Photo-Optical Instrumentation Engineers (SPIE) Conference,
  Vol. 733, Society of Photo-Optical Instrumentation Engineers (SPIE)
  Conference Series, ed. {E.-E.~Koch \& G.~Schmahl}, 519

\bibitem[{{Predehl} \& {Prieto}(2001)}]{predehl01}
{Predehl}, P. \& {Prieto}, A. 2001, ArXiv:astro-ph/0109542

\bibitem[{Roberts {et~al.}(2005)Roberts, Brogan, Gaensler, Hessels, NG, \&
  Romani}]{roberts05}
Roberts, M., Brogan, C., Gaensler, B., {et~al.} 2005, Astrophysics and Space
  Science, 297, 93

\bibitem[{{Rosado} {et~al.}(1996){Rosado}, {Ambrocio-Cruz}, {Le Coarer}, \&
  {Marcelin}}]{rosado1996}
{Rosado}, M., {Ambrocio-Cruz}, P., {Le Coarer}, E., \& {Marcelin}, M. 1996,
  \aap, 315, 243

\bibitem[{{Stappers} {et~al.}(2003){Stappers}, {Gaensler}, {Kaspi}, {van der
  Klis}, \& {Lewin}}]{stappers2003}
{Stappers}, B.~W., {Gaensler}, B.~M., {Kaspi}, V.~M., {van der Klis}, M., \&
  {Lewin}, W.~H.~G. 2003, Science, 299, 1372

\bibitem[{{Tanaka} {et~al.}(1994){Tanaka}, {Inoue}, \& {Holt}}]{tanaka94}
{Tanaka}, Y., {Inoue}, H., \& {Holt}, S.~S. 1994, PASJ, 46, L37

\bibitem[{{Truemper}(1982)}]{trumper82}
{Truemper}, J. 1982, Advances in Space Research, 2, 241

\bibitem[{{Ubertini} {et~al.}(2003){Ubertini}, {Lebrun}, {Di Cocco}, {Bazzano},
  {Bird}, {Broenstad}, {Goldwurm}, {La Rosa}, {Labanti}, {Laurent}, {Mirabel},
  {Quadrini}, {Ramsey}, {Reglero}, {Sabau}, {Sacco}, {Staubert}, {Vigroux},
  {Weisskopf}, \& {Zdziarski}}]{ubertini03}
{Ubertini}, P., {Lebrun}, F., {Di Cocco}, G., {et~al.} 2003, \aap, 411, L131

\bibitem[{Vink \& Bamba(2009)}]{vink2009}
Vink, J. \& Bamba, A. 2009, ApJ, 707, L148

\bibitem[{{Voges} {et~al.}(1999){Voges}, {Aschenbach}, {Boller},
  {Br{\"a}uninger}, {Briel}, {Burkert}, {Dennerl}, {Englhauser}, {Gruber},
  {Haberl}, {Hartner}, {Hasinger}, {K{\"u}rster}, {Pfeffermann}, {Pietsch},
  {Predehl}, {Rosso}, {Schmitt}, {Tr{\"u}mper}, \& {Zimmermann}}]{voges99}
{Voges}, W., {Aschenbach}, B., {Boller}, T., {et~al.} 1999, A\&A, 349, 389

\bibitem[{Walter {et~al.}(2010)Walter, Rohlfs, Meharga, Binko, Morisset, Beck,
  Produit, Pavan, Savchenko, Ferrigno, Frankowski, \& Bordas}]{heavens}
Walter, R., Rohlfs, R., Meharga, M., {et~al.} 2010, PoS(INTEGRAL 2010), 162

\bibitem[{{Watson} {et~al.}(2009){Watson}, {Schr{\"o}der}, {Fyfe}, {Page},
  {Lamer}, {Mateos}, {Pye}, {Sakano}, {Rosen}, {Ballet}, {Barcons}, {Barret},
  {Boller}, {Brunner}, {Brusa}, {Caccianiga}, {Carrera}, {Ceballos}, {Della
  Ceca}, {Denby}, {Denkinson}, {Dupuy}, {Farrell}, {Fraschetti}, {Freyberg},
  {Guillout}, {Hambaryan}, {Maccacaro}, {Mathiesen}, {McMahon}, {Michel},
  {Motch}, {Osborne}, {Page}, {Pakull}, {Pietsch}, {Saxton}, {Schwope},
  {Severgnini}, {Simpson}, {Sironi}, {Stewart}, {Stewart}, {Stobbart}, {Tedds},
  {Warwick}, {Webb}, {West}, {Worrall}, \& {Yuan}}]{watson09}
{Watson}, M.~G., {Schr{\"o}der}, A.~C., {Fyfe}, D., {et~al.} 2009, A\&A, 493,
  339

\bibitem[{{Winkler} {et~al.}(2003){Winkler}, {Courvoisier}, {Di Cocco},
  {Gehrels}, {Gim{\'e}nez}, {Grebenev}, {Hermsen}, {Mas-Hesse}, {Lebrun},
  {Lund}, {Palumbo}, {Paul}, {Roques}, {Schnopper}, {Sch{\"o}nfelder},
  {Sunyaev}, {Teegarden}, {Ubertini}, {Vedrenne}, \& {Dean}}]{winkler03}
{Winkler}, C., {Courvoisier}, T., {Di Cocco}, G., {et~al.} 2003, \aap, 411, L1

\end{thebibliography}

\end{document}